\documentclass[twocolumn,amsmath,amssymb]{revtex4}

\usepackage{graphicx}% Include figure files
\usepackage{dcolumn}% Align table columns on decimal point
\usepackage{bm}% bold math

\begin{document}
%\firstpage{1}

%\title[statistical framework for detecting DGEs]
%{A statistical framework for the design of microarray experiments and effective detection of differential gene expression}
%\author[Zhang and Gant]{Shu-Dong Zhang\,$^{\rm a}$, Timothy W. Gant\,$^{\rm a}$}
%\address{$^{\rm a}$MRC Toxicology Unit, Hodgkin Building, Lancaster Road,
%University of Leicester, Leicester, UK}

\title{A statistical framework for the design of microarray
experiments and effective detection of differential gene
expression}
\author{\small Shu-Dong Zhang and Timothy W. Gant\\
\small MRC Toxicology Unit, Hodgkin Building, Lancaster Road,
University of Leicester, Leicester, UK }

\begin{abstract}

%\section{Motivation:}
Microarray experiments generate a high data volume. However, often
due to financial or experimental considerations, e.g. lack of
sample, there is little or no replication of the experiments or
hybridizations. These factors combined with the intrinsic
variability associated with the measurement of gene expression can
result in an unsatisfactory detection rate of differential gene
expression (DGE). Our motivation was to provide an easy to use
measure of the success rate of DGE detection that could find
routine use in the design of microarray experiments or in
post-experiment assessment.

%\section{Results:}
In this study, we address the problem of both random errors and
systematic biases in microarray experimentation. We propose a
mathematical model for the measured data in microarray experiments
and on the basis of this model present a t-based statistical
procedure to determine DGE. We have derived a formula to determine
the success rate of DGE detection that takes into account the
number of microarrays, the number of genes, the magnitude of DGE,
and the variance from biological and technical sources. The
formula and look-up tables based on the formula, can be used to
assist in the design of microarray experiments. We also propose an
ad hoc method for estimating the fraction of non-differentially
expressed genes within a set of genes being tested. This will help
to increase the power of DGE detection.

%\section{Availability:}
The functions to calculate the success rate of DGE detection have
been implemented as a Java application, which is accessible at

http://www.le.ac.uk/mrctox/microarray\_lab/Microarray\_Softwares/Microarray\_Softwares.htm.
Supplementary information at ftp://alcyone.mrc.le.ac.uk/
Pub/twg1/BioInf03-0661suppl.pdf

%http://www.le.ac.uk/mrctox/microarray_lab/Microarray_Softwares/Microarray_Softwares.htm
%\section{Contact:} \href{sdz1@le.ac.uk;  twg1@le.ac.uk}{sdz1@le.ac.uk;  twg1@le.ac.uk}
\end{abstract}

\maketitle

\section{Introduction} \label{sec-introduction} Whole genome sequencing and
the related development of microarrays have given researchers
unprecedented power to simultaneously determine the expressions of
many thousands of genes \cite{Baldi&Hatfield2002}. However, a
statistical challenge facing microarray analysis is to identify
differential gene expression (DGE) with a high rate of success and
low rate of false positives. Such a method is required because of
the number of gene expressions being simultaneously determined,
and the variation associated with each can give rise to an
unacceptably large number of false positives or low successful
detection rate. The variations associated with gene expression
experiments can be categorized into two sets. First, there are
inter-individual differences between members of a population, thus
sufficient biological individuals should be included in the
experiments in order to account for the biological variation.
Second, there are always technical errors arising from the
experimental procedure, which may be further sub-categorized into
random errors and systematic biases. Unlike random errors, which
can be reduced by making multiple measurements, systematic biases
cannot be reduced by simply doing more measurements, correct
experimental designs must be employed to negate them.

One of the most serious sources of systematic bias in microarray
experiments (for dual label hybridizations) is the imbalance in
the measured fluorescence intensities between the two fluorescent
channels \cite{Dudoit-etal2002,Yang-etal2002,Dobbin-etal2003}. A
manifestation of this systematic bias is that when two identical
mRNA samples are labelled with different fluorescent dyes and
hybridized to the same microarray slide, one channel has a higher
average fluorescence level than the other. To complicate matters
further the imbalance of the two channels is not uniform, but
varies from feature to feature. A feature is the area of
fluorescence on a microarray corresponding to one gene and where
hybridization of the labelled nucleic acids derived from this gene
has taken place \cite{Cheung-etal1999}. To correct the labelling
dye imbalance, different methods of normalizing the microarray
data by adjusting the measured fluorescence levels have been
proposed \cite{Yang-etal2002,Gant&Zhang2002,Quackenbush2002}.
These methods can be roughly classified into two categories.
First, global normalization, in which the fluorescence levels of
all the features are globally (uniformly) adjusted (by shifting or
re-scaling) to fulfill some assumptions about the relative
expressions of the genes, e.g. most genes are not differentially
expressed between the two samples \cite{Gant&Zhang2002}. However,
because global normalization adjusts the fluorescence levels of
all features uniformly, it cannot account for the different
magnitudes of imbalances from feature to feature, so a second type
of normalization method is often employed to take account of this
variation. This normalization method adjusts the fluorescence
level according to some local properties of the feature spot, e.g.
the overall brightness of the spot \cite{Yang-etal2002}, and
usually involves fitting the measured data with a non-linear
smoothed curve. The fluorescence level is then adjusted according
the smoothed curve, which is assumed to describe the dependence of
the imbalance on spot fluorescence intensity. But the fluorescence
imbalances between the two channels are more complicated than can
be described by a smoothed curve. Due to irregular intrinsic
fluorescence of the microarray slide and possibly some
gene-specific effect \cite{Tseng-etal-2001,Zhou-etal-2002}, it is
unlikely that the fluorescence imbalance can be corrected for all
features by the intensity-dependent normalization. To correct the
fluorescence imbalance of each feature a simple method is to
reverse the labelling dyes when hybridizing some microarrays. Kerr
{\em et al} \cite{Kerr-etal-2000} first proposed an ANOVA model
for microarray data, and showed that ANOVA methods can be used to
normalize the data and estimate real changes in gene expression.
Taking biological variations into account, Dobbin {\em et al}
\cite{Dobbin-etal2003} have addressed the problem of statistical
design of reverse dye microarrays to minimize variance with a
given number of microarray slides. We have taken the analysis
further to address the problem of identifying DGEs with a desired
detection power and controlled number of false positives. A model
and statistical testing procedure are presented in the following
sections to assist research workers in the selection of an
appropriate number of microarrays for an experiment in order to
achieve the desired detection power, or alternatively in assessing
the detection power achievable when the experiment has been done.

\section{The model}
The experimental situation analyzed here is one where there are
two sample groups. One of the groups might have been subjected to
an event such as chemical exposure, the other being a suitable
control, or the two groups might be normal and tumor tissues or
different organs. For convenience the two groups will be
designated as the {\em treated} and {\em control} groups.

In cells, the amount of mRNA corresponding to a particular gene is
taken to correspond to the expression level of that gene. A
microarray is a means to translate the level of mRNA for many
genes, which cannot be measured directly, into fluorescence that
can be measured directly. The model presented in this paper is
designed for experiments where each gene is spotted only once on
each microarray, and each individual sample is hybridized only
once using one microarray. For the purpose of introduction
consider one single feature spot on the microarray. We assume that
the log-intensity fluorescence of this feature takes additive
contributions from the following sources: the amount of
corresponding mRNA in the biological sample, an effect from the
quality of the feature spot, an effect from the labelling
fluorochrome (including the efficiency of labelling with the
fluorochrome, and possible pre-existing intrinsic fluorescence in
favor of this fluorochrome), and the random measurement error.
Therefore we have the following model
\begin{equation}
G_{v,i,s,c}=I_{v,i}+A_s+D_c+\epsilon _{v,i,s,c}, \label{eq-G_visc}
\end{equation}
where $G_{v,i,s,c}$ is the log-intensity (base 2 logarithms are
utilized throughout this paper) of fluorescence of the feature
spot;

$I_{v,i}$ is the expression level of the gene in the $i$th
individual sample of group $v$ ($v=t$ for the treated group, or
$v=c$ for the control group, and $i=1\cdots n$ where $n$ is the
number of individuals in each group). $I_{v,i}$ is assumed to be
independently and normally distributed with a mean $E_v$ and a
variance $\sigma _v^2$, denoted by $I_{v,i}\sim
N(E_v,\sigma_v^2)$;

$A_s$ is the effect of feature spot quality, which is assumed to
be fixed for microarray slide $s$ and independent of fluorescent
label used;

$D_c$ is the effect of the fluorescent label $c$ ($c=g$ for green
dye, and $c=r$ for red dye), which is assumed to be fixed with
label $c$ and independent of microarray slide;

$\epsilon _{v,i,s,c}$ is the random error term which is assumed to
be independently and normally distributed with a mean $0$ and a
variance $\sigma _{\epsilon}^2$, denoted by $\epsilon _{v,i,s,c}
\sim N(0,\sigma _{\epsilon}^2)$.

Note that for each of the features on the microarray the
log-intensity is described in the same form Eq.(\ref{eq-G_visc}).
Although the equations are in the same form for each feature the
actual values of $E_v$, $\sigma_v^2$, $A_s$, $D_c$, $\sigma
_{\epsilon}^2$ will be feature dependent.

$E_v$ is the mean expression level of the gene in the sample group
$v$, so in comparing a gene's expression between the treated and
control groups, the quantity of interest is $E_t-E_c$, the
magnitude of differential expression. The effects of feature spot
quality $A_s$ and fluorescent dye $D_c$ are not of interest and
therefore need to be eliminated by a suitable experimental design.

\section{Experimental setup}
Let's introduce the notation
$(c_j,t_i)$ to represent a microarray as a result of the following
hybridization: individual sample $c_j$ labelled with {\em green}
dye and individual sample $t_i$ labelled with {\em red} dye. Here
$c$ and $t$ indicate the sample group while the subscripts index
different individuals in each group. As a convention, the first
sample in the parenthesis is always labelled with green dye and
the second with red dye.

Consider the microarray $(c_j,t_i)$. Here an individual $j$ is
taken from the control group ($v=c$) and an individual $i$ from
the treated group ($v=t$). RNA is extracted from both and
converted to labelled cDNA using fluorescent labels {\em green}
and {\em red} respectively. These are then simultaneously
hybridized to the microarray $a$. This method of labelling
(control sample with {\em green} and treated sample with {\em
red}) is referred to as {\em forward labelling}. As a result of
this experiment we can derive from Eq.(\ref{eq-G_visc})
\begin{eqnarray}
G_{c,j,a,g}=I_{c,j}+A_a+D_{g}+\epsilon _{c,j,a,g},\nonumber \\
%\label{eq-G_control_green}
G_{t,i,a,r}=I_{t,i}+A_a+D_{r}+\epsilon _{t,i,a,r}. \nonumber
%\label{eq-G_treat_red}
\end{eqnarray}
The difference $F_a$ between the two fluorescence log-intensities
is therefore
\begin{eqnarray}
F_a=G_{t,i,a,r}-G_{c,j,a,g} =I_{t,i}-I_{c,j} \nonumber
\\+D_{r}-D_{g}
 +\epsilon
_{t,i,a,r}-\epsilon_{c,j,a,g}, \label{eq-forward}
\end{eqnarray}
and $F_a$ is normally distributed with an expected value (mean)
$E_t-E_c+D_{r}-D_{g}$ and a variance $\sigma _t^2+\sigma
_c^2+2\sigma _{\epsilon} ^2$. Note that taking the difference of
$G_{c,j,a,g}$ and $G_{t,i,a,r}$ causes the spot effect $A_a$ to be
cancelled out and it does not therefore contribute to $F_a$.
However, there is still the labelling fluor effect $D_{r}-D_{g}$
to consider. To eliminate this effect microarrays with reverse
labelling are required.

Consider the microarray $(t_{i'},c_{j'})$, where another two
individuals, $i'$ from the treated and $j'$ from the control
groups are hybridized to another microarray $b$. On this occasion
the individual from the control group $j'$ is labelled with $red$
and the individual from the treated group $i'$ with $green$. This
method of labelling (control sample with $red$ and treated sample
with $green$) is referred to as {\em reverse labelling}. From this
microarray $b$ we get:
\begin{eqnarray}
G_{t,i',b,g}=I_{t,i'}+A_b+D_{g}+\epsilon _{t,i',b,g},\nonumber\\
%\label{eq-G_treat_green}
G_{c,j',b,r}=I_{c,j'}+A_b+D_{r}+\epsilon _{c,j',b,r}.\nonumber
%\label{eq-G_control_red}
\end{eqnarray}
The difference of the two log-intensities is
\begin{eqnarray}
B_b=G_{t,i',b,g}-G_{c,j',b,r} =I_{ti'}-I_{cj'}\nonumber
\\+D_{g}-D_{r}
 +\epsilon
_{t,i',b,g}-\epsilon_{c,j',b,r},  \label{eq-reverse}
\end{eqnarray}
and $B_b$ is normally distributed with an expected value
$E_t-E_c+D_{g}-D_{r}$ and a variance $\sigma _t^2+\sigma
_c^2+2\sigma _{\epsilon} ^2$.

The quantity $F_a$ (or $B_b$) is the difference of two
log-intensities and is therefore equivalent to the logarithm of
the ratio of two intensities. Thus $F_a$ (or $B_b$) is often
called the {\em log-ratio} of a gene.  The variance of the
log-ratio of a gene, $\sigma _T^2=\sigma _c^2+\sigma _t^2+2\sigma
_{\epsilon} ^2$, is the sum of the biological variance of the
control $\sigma _c^2$, the treated $\sigma _t^2$, and the
measurement variances associated with them $2\sigma _{\epsilon}
^2$. Hereafter $\sigma_T^2$ is referred to as the {\em total
variance} of the log-ratio of the gene.

From Eqs.(\ref{eq-forward}) and (\ref{eq-reverse}) it is clear
that by combining measurements from both forward and reverse
labelled microarrays, it is possible to eliminate the fluorescent
label bias. One simple way of doing this is to take the average of
Eqs.(\ref{eq-forward}) and (\ref{eq-reverse}). The expected value
of this average is then $E_t-E_c$, which is the quantity of
interest. The above arguments therefore show that to eliminate the
spot effect $A_s$, we need to hybridize the control and treated
samples onto the same microarray slide. To cancel out the
fluorescent label effect $D_c$ we need to do both forward labelled
and reverse labelled microarrays. A general formalism is presented
in the following sections to deal with situations where the number
of forward labelled microarrays and the number of reverse labelled
microarrays are not necessarily the same.

We will consider the following experiment:
\begin{eqnarray}
(c_1,t_1), (c_2,t_2), \cdots,
(c_{n_f-1},t_{n_f-1}),(c_{n_f},t_{n_f}) \nonumber \\
(t_{n_f+1},c_{n_f+1}), (t_{n_f+2},c_{n_f+2}), \cdots,
(t_{n_f+n_r},c_{n_f+n_r}). \nonumber
\end{eqnarray}
In this experiment there are in total $n_f+n_r$ microarrays, $n_f$
of them are forward labelled, and the rest $n_r$ are reverse
labelled. In relation to similar studies by other
authors\cite{Jin-etal2001,Kerr-etal2002,Callow-etal2000} using
replicated microarrays, this study focuses on a special case of
microarray experiment designs, i.e., direct comparison between two
groups with biological but no technical replicates in each group.
It is a special case of the balanced block design as described by
\cite{Dobbin&Simon2002}. They have showed that the balanced block
design is the most efficient experimental setup when comparing two
classes with a given number of microarrays. The limitation of this
experimental setup, as Dobbin and Simon pointed out for the
balanced block design, is that it is not suitable for clustering
analysis.

\section{ Detecting DGE\lowercase{s}}
\label{sec-method}
\subsection{Hypothesis test}
For each gene printed on the microarrays, we want perform a
statistical test to determine whether this gene is differentially
expressed to a significant degree in the treated group compared to
the control group. The null hypothesis is that the gene has the
same expression level in the two groups:
\begin{eqnarray}
\mbox{Null hypothesis } H_0:  E_c=E_t \label{eq-null}\\
\mbox{Alternative hypothesis } H_1:   E_c\neq
E_t\label{eq-alternative}
\end{eqnarray}
From each of the $n_f$ forward labelled microarrays an intra-array
log-ratio $F_i$ between the treated sample and the control sample
is obtained, and similarly from each of the $n_r$ reverse labelled
microarrays a log-ratio $B_j$. Each $F_i$ has an expected value
$E_t-E_c+D_{r}-D_{g}$, so the average $\overline{F}=\sum
_{i=1}^{n_f}F_i/n_f$ has the same expected value. Similarly the
average $\overline{B}=\sum _{j=1}^{n_r}B_j/n_r$ has an expected
value $E_t-E_c-D_{r}+D_{g}$. Averaging $\overline{F}$ and
$\overline{B}$ gives
\begin{equation}
R=\frac{\overline{F}+\overline{B}}{2}={1\over 2n_f}\sum
_{i=1}^{n_f}F_i+{1\over 2n_r}\sum _{j=1}^{n_r}B_j, \label{eq-R}
\end{equation}
which will have an expected value $E_t-E_c$, so $R$ is an unbiased
estimator of our  quantity of interest. Also $R$ is normally
distributed with a variance
\begin{equation}
\sigma_R^2
%=\frac{\sigma _c^2+\sigma _t^2+2\sigma _{\epsilon}^2}{4}\left(\frac{1}{n_f}+\frac{1}{n_r}\right)
=\frac{\sigma_T^2}{4}\left(\frac{1}{n_f}+\frac{1}{n_r}\right).
\label{eq-sigma_R^2}
\end{equation}
When the total number of microarrays $n_f+n_r$ is fixed, the
variance of $R$ is minimized at $n_f=n_r$, so whenever possible,
equal numbers of forward and reverse labelled microarrays should
be combined. The variances $\sigma _c^2$, $\sigma _t^2$, and
$\sigma_{\epsilon}^2$ are unknowns, but fortunately there is no
need to estimate them individually. For the purposes of
identifying differential gene expression, estimating $\sigma_T^2$
as a whole is sufficient and $\sigma_T^2$ can be estimated using
its un-biased estimator
\begin{equation}
s^2={1\over n_f+n_r-2}\left[\sum _{i=1}^{n_f}(F_i-\overline{F})^2+
\sum _{j=1}^{n_r}(B_j-\overline{B})^2\right] \label{eq-s^2}
\end{equation}
and $(n_f+n_r-2)s^2/\sigma_T^2$ will follow the $\chi^2$
distribution with $n_f+n_r-2$ degrees of freedom, independent of
$\overline{F}$ and $\overline{B}$\cite{Brownlee1965}, thus $s^2$
is independent of $R$. Note that in order to estimate $E_t-E_c$
and $\sigma_T^2$ properly it is necessary that $n_f \ge 1$,
$n_r\ge 1$, and $n_f+n_r> 2$. In other words there  must be at
least one forward and one reverse labelled microarray, and at
least three microarrays in total. It is then apparent that
\begin{equation}
t={R-(E_t-E_c)\over
s\sqrt{\frac{1}{4}\left(\frac{1}{n_f}+\frac{1}{n_r}\right)}}
\label{eq-t}
\end{equation}
is distributed as the Student's $t$ distribution with $n_f+n_r-2$
degrees of freedom. In testing the null hypothesis
Eq.(\ref{eq-null}), we insert $E_t=E_c$ into Eq.(\ref{eq-t}) and
thus our test statistic $t_0$ is defined as,
\begin{equation}
t_0={R\over
s\sqrt{\frac{1}{4}\left(\frac{1}{n_f}+\frac{1}{n_r}\right)}}
\label{eq-t_0}
\end{equation}
Note that there is now no unknown quantity in Eq.(\ref{eq-t_0}).
Under the null hypothesis that $E_t=E_c$, $t_0$ follows the
Student's distribution with $n_f+n_r-2$ degrees of freedom. Based
on the value of $t_0$ the p-value of the test can be calculated.
If the p-value calculated is larger than some pre-set threshold
$P_{th}$, the null hypothesis is accepted that the gene has the
same level of expression in both the control and treated groups.
If the calculated p-value is smaller than the threshold $P_{th}$,
it is declared that the test for this gene is {\em positive}, in
the sense that its expression level in the treated group is
different from that in the control group. Then depending on the
sign of $t_0$ the gene is either designated as up ($t_0>0$) or
down regulated ($t_0<0$).

\subsection{Setting the threshold p-value}
\label{sec-setting_threshold} A t test is performed for each gene,
which is then declared as differentially expressed, or not,
according to the above criteria. By adjusting the value of
threshold $P_{th}$ a control can be exerted on  the number of
false DGE calls made. By definition, p-value is the probability of
observing a value of the statistic as extreme or more extreme than
the observed value, under the condition that the null hypothesis
is true. For each gene whose null hypothesis is true (we call each
such gene a {\em null gene}), its p-value is uniformly distributed
in $(0,1)$. Therefore the probability that a null gene's p-value
is smaller than $P_{th}$ is just $P_{th}$. Suppose that in a total
number $N$ genes, $N_0$ are null genes. When every gene on the
microarray is tested, the number of false DGE calls $O_{fp}$ will
has an expected value $N_0P_{th}$. So if one decides to tolerate
an expected number $N_{fp}$ false DGEs the threshold p-value
should be set at $P_{th}=N_{fp}/N_0$. However, in reality only $N$
is known and not $N_0$ and therefore, it is necessary to make an
estimation of $N_0$ or $N_0/N$. Some methods for estimating
$N_0/N$ are discussed in Sec. \ref{sec-estimatingN0/N}.

Once the threshold value $P_{th}$ is set, the ability to detect
genuine DGE, i.e. a gene with $E_t\neq E_c$, depends on the
following factors: the magnitude of differential expression
$E_t-E_c$, the total variance in one microarray experiment $\sigma
_T^2$, and the number of forward and reverse labelled microarrays.
Among these factors, the ones over which experimental control is
exercised are $n_f$ and $n_r$. In general the larger $n_f$ and
$n_r$, the more powerful will be the statistical testing. The key
question is therefore, how many forward and reverse labelled
microarrays are required in order to achieve a desired power of
DGE detection with control on the number of false DGE calls? Based
on the standard normal Z test, several authors have presented
results on calculating the number of microarrays needed to achieve
given statistical power while controlling false positive rate
\cite{Wernisch2002,Dobbin-etal2003}. These results would be
applicable if we knew $\sigma _T^2$ for each gene. In reality
though the variances cannot be assumed known, and more often than
not, the number of microarrays used to estimate the variances is
rather small. It is therefore necessary to use $t$-based test
rather than the standard normal test. Other authors have also
presented approximate formulas
\cite{McShane-etal2003,Simon-etal2004} for calculating the power
of the traditional two-sample $t$ test with equal variance. In
this paper we present an exact formula for calculating the power
of the t-based statistical test developed here.

\subsection{Determination of the threshold t-value}
When the numbers of forward and reverse labelled microarrays are
given, setting $P_{th}$ is equivalent to setting a threshold, say
$|\xi|$, for the statistics $t_0$ defined in Eq.(\ref{eq-t_0}).
With this threshold t-value, our criteria for claiming a DGE is as
follows: If $t_0>|\xi|$, the gene is claimed as up-regulated
($E_t-E_c>0$); if $t_0<-|\xi|$, it is claimed as down-regulated
($E_t-E_c<0$). So the rate at which false positive claims are made
is
\begin{eqnarray}
P_{th}=\int _{-\infty}^{-|\xi|}\rho _{n_f+n_r-2}(t_0)dt_0
+\int_{|\xi|}^{\infty}\rho _{n_f+n_r-2}(t_0)dt_0 \nonumber \\
=2\int _{-\infty}^{-|\xi|}\rho _{n_f+n_r-2}(t_0)dt_0
=2T_{n_f+n_r-2}(-|\xi|), \label{eq-P_th}
\end{eqnarray}
where $\rho_r(x)$ is the probability density function (PDF) of the
Student's distribution with $r$ degrees of freedom, and $T_r(.)$
is the cumulative probability distribution function (CDF) for the
Student's t distribution. It is therefore apparent that the
threshold t-value $|\xi|$ can be obtained by solving the equation
$2T_{n_f+n_r-2}(-|\xi|)=P_{th}$ with a given false positive rate
$P_{th}$.

\subsection{Successful detection rate}
\label{sec-S} The successful detection rate is the rate at which
DGE is correctly identified (either up-regulated or
down-regulated). If a gene has $E_t-E_c=\mu>0$, the successful
detection rate for this gene is the probability that $t_0>|\xi|$
is observed. On the other hand, if a gene has $E_t-E_c=\mu <0$,
the successful detection rate equals the probability that
$t_0<-|\xi|$ is observed. It can be shown (see supplementary
information I) that in both cases, the rate at which the genes
behavior is correctly identified, i.e. $\mu>0$ or $\mu<0$, can be
described by the following equation
\begin{eqnarray}
S\left(n_f,n_r,\frac{|\mu|}{\sigma _T},|\xi|\right)
=\int _0^{\infty}p_{n_f+n_r-2}(Y)\times  \nonumber \\
\Phi\left[-|\xi|\sqrt{Y\over n_f+n_r-2}+2\left(\frac{|\mu|}{\sigma
_T}\right)\sqrt{n_fn_r\over n_f+n_r}\right]dY,
 \label{eq-S}
\end{eqnarray}
where $p_r(Y)$ is the PDF for the $\chi ^2$ distribution with $r$
degrees of freedom, and $\Phi(.)$ is the CDF for the standard
normal distribution.

Therefore the successful detection rate $S$ is a function of
$n_f$, $n_r$, $|\mu|/\sigma _T$, and $|\xi|$, where $|\xi|$ can be
obtained by solving Eq.(\ref{eq-P_th}) at a given $P_{th}$.
Eventually, $S$ is a function of $P_{th}$, $n_f$, $n_r$, and
$|\mu|/\sigma _T$.

\subsection{Usage of the $S$ function}
We have implemented the calculation of the $S$ function as a Java
application, which is accessible through the URL given in the
abstract. Two look-up tables also are provided in the
supplementary for some typical results of $S$ for quick reference.
Experiment designers can use these to find the value of $S$ at
given parameters $n_f$, $n_r$, $|\mu|/\sigma_T$, and $P_{th}$,
thus get some general idea of what percentage of truly DGEs can be
detected by their experimental design.

The applicability of the $S$ function can be seen from two
perspectives. First, for the user who has not carried out any
microarray experiments on their system before, the total variances
($\sigma_T^2$) will be completely unknown. In this situation the
$S$ function can serve as a post-experiment assessment to inform
the user of the detection rate in their experiment based on the
observed values of $R$ and $s^2$ from the measurements. For
example, $3$ forward and $3$ reverse labelled microarrays, with
$5000$ genes printed on each microarray, were used in a
experiment. The tolerance for false positives is set at
$N_{fp}=2$, and for simplicity the threshold p-value is set as
$P_{th}=2/5000$. If most genes have an $s^2$ around $1$, then the
typical value of $\sigma_T^2$ for the set of genes is $1$. We can
now ask: for genes with two-fold differential expression and
typical variance, what percentage of them can be correctly
detected by this experiment? Remembering that a two-fold
differential expression corresponds to $\mu=E_t-E_c=1$ or
$\mu=E_t-E_c=-1$, we have $|\mu|=1$ and $\sigma_T=1$. Using the
$S$ calculator or the look-up tables (Supplementary Table I) we
find that the successful detection rate for $n_f=3$, $n_r=3$,
$P_{th}=2/5000$, and $|\mu|/\sigma_T=1$ is $9.08\times 10^{-3}$,
which means that in this experiment only $0.908\%$ of genes with
two-fold DGE and with typical variance $1$ can be detected, the
remaining $99\%$ are missed. If the same question was asked about
genes with four-fold DGE and one decides to tolerate $N_{fp}=8$
false positive claims and the threshold is set at $P_{th}=8/5000$,
then the successful detection rate for $P_{th}=8/5000$, $n_f=3$,
$n_r=3$, and $|\mu|/\sigma_T=2$ is $0.217$, which means $21.7\%$
of them are successfully detected. If the detection rate is
unsatisfactory, then more forward and reverse microarray datasets
need to be added.

Second, if there is some general knowledge of total variance from
previous experiments or other sources, then a target for the
detection rate can be set. In this case, the $S$ function will
assist in the determination of how many forward and reverse
microarrays are required in the experiment. For example, if from
previous experience we know that the typical value of the total
variance for the set of genes under consideration is
$\sigma_T^2=0.25$, which gives $\sigma_T=0.5$; A microarray
experiment is now designed to identify DGEs between the treated
and the control with a tolerance of $8$ false positive claims out
of $5000$ genes being tested with $P_{th}=8/5000$ for simplicity;
The pre-set target is that after this experiment no less than
$60\%$ of genes with two-fold DGE and with typical variance should
be detected; How many forward and reverse labelled microarrays are
needed? As before, two-fold DGE corresponds to $|\mu|=1$, so one
has $|\mu|/\sigma_T=2$. Using the look-up tables (Supplementary
Table II, in the $|\mu|/\sigma_T=2$ panel and $P_{th}=8/5000$
column), one finds that the row $n_f=n_r=4$ gives a detection rate
$S=0.605$ which is closest to meet the target. Therefore $4$
forward and $4$ reverse labelled microarrays are required in this
experiment.

\section{Controlling false positives}
\subsection{Procedures}
In this section, we explore further on how to effectively control
false positives in a multiple test situation. Generally speaking,
all different multiple-testing methods eventually amount to
effectively setting a threshold p-value, and then rejecting all
the null hypothesis with p-value below this threshold. For
example, the classical Bonferroni multiple-testing procedure
controls family-wise error rate at $\alpha$ by setting the
threshold $P_{th}=\alpha /N$, where $N$ is the total number of
hypothesis tested. In this study, we aim to control the number
false positives such that the expectation of $O_{fp}$ equals
$N_{fp}$, our pre-set target. As discussed in Section
\ref{sec-setting_threshold}, to achieve this, we should set
$P_{th}=N_{fp}/N_0$, which requires an estimation of $N_0$ or
$N_0/N$, the fraction of null genes in the set.

We present three procedures here for setting $P_{th}$ to control
false positives:

Procedure A: Suppose we have made an estimation of $N_0/N$ as $c$,
then set $P_{th}=N_{fp}/(cN)$. The method for calculating $c$ will
be discussed below.

Procedure B: Set $P_{th}=N_{fp}/N$. This can be seen as using
$c=1$ as the crudest estimation of $N_0/N$.

Procedure C: Suppose genes are sorted by their ascending p-values,
so that $p_1\le p_2 \le p_3 \le \cdots \le p_N$, where $p_i$ is
the p-value for gene $i$. Set $P_{th}=p_{i^*}$, where $i^*$ is the
largest index satisfying $p_i[N-i+\min (i,N_{fp})]\leq N_{fp}$.
This can be seen as estimating $N_0/N$ by $c=[N-i^*+\min
(i^*,N_{fp})]/N$. The idea behind this is that if gene $i^*$ and
all genes indexed below it are to be declared DGEs, these genes
should not contribute to the fraction of null genes. Thus this
represents some improvement over the crudest estimation $c=1$.

We have performed simulations to compare the performances of the
three procedures. Procedure A allows us to achieve the highest
rate of DGE detection among the three, and the observed false
positives $O_{fp}$ matches our preset target $N_{fp}$
statistically. Procedure B does not estimate $N_0/N$ effectively,
and it is the most conservative procedure. So Procedure A is
recommended over C and B (See Supplementary for details on
simulation procedures and data).

Benjamini and Hochberg \cite{Benjamini&Hochberg1995} proposed the
FDR approach to control the false discovery rate (FDR) at $q$ by
setting $P_{th}=i^*q/N$, where $i^*$ is the largest index
satisfying $p_i\leq iq/N$. The false discovery rate was defined as
the expectation of the ratio of false to total positives, i.e.,
$q\equiv E( O_{fp}/i^*)$. When the FDR procedure controls false
discovery rate at $q$, the observed false discovery rate
$O_{fp}/i^*$ should have value around $q$, i.e., $q\approx
O_{fp}/i^*$, which gives $P_{th}=i^*q/N\approx O_{fp}/N$. The
expectation of the threshold p-value under the FDR procedure is
therefore $E(P_{th})\approx E(O_{fp}/N)=N_{fp}/N$. It is thus
clear that the FDR procedure of \cite{Benjamini&Hochberg1995} is
on average equivalent to Procedure B in this section.

\subsection{Estimating $N_0/N$}
\label{sec-estimatingN0/N}Pounds and Morris
\cite{Pounds&Morris2003} recently proposed the use of a
beta-uniform mixture (BUM) function to approximate the
distribution of p-values from a set of genes tested, and estimate
the fraction of null genes in the set. Here we propose another
method to estimate $N_0/N$, which does not requires the BUM form
of distribution of p-values. The aim was to achieve a more
accurate estimation of the fraction of null genes. As in
\cite{Pounds&Morris2003}, we wanted to extract a uniform density
from the observed distribution of p-values. To achieve this, the
genes were first sorted by their ascending p-values, so that
$p_1\le p_2 \le p_3 \le \cdots \le p_N$, where $p_i$ is the
p-value for gene $i$. Then an empirical cumulative distribution of
p-values can be easily obtained by plotting $i/N$ versus $p_i$.
The idea was to find a straight line tangent to the cumulative
distribution curve with minimum slope. Taking into account that
the cumulative distribution curve is a non-decreasing function
ending at the point $(1.0,1.0)$, the minimum slope was found as
follows. Each point $(p_i,i/N)$ on the cumulative distribution
plot was connected with the ending point $(1.0,1.0)$ with a
straight line, and the slope of the line calculated as
$c_i=(1.0-i/N)/(1.0-p_i)$. Then the minimum of $c_i$ at a given
range of p-value, say $P_l\le p_i \le P_u$, was found
\begin{equation}
c_{min}=\min_i ( c_i \mid P_l\le p_i \le P_u ). \label{eq-cmin}
\end{equation}
$c_{min}$ can be used as our estimation of the fraction of null
genes in the set.

We have carried out simulations to test the performance of
Eq.(\ref{eq-cmin}), and found that it tends to underestimate the
true value of $N_0/N$. Instead, using median slope as the
estimation of $N_0/N$ gives more accurate results than the minimum
slope. We thus use the following equation to estimate the fraction
of null genes
\begin{equation}
c_{mid}=\mbox{median} ( c_i \mid P_l\le p_i \le P_u ).
\label{eq-cmedian}
\end{equation}

In a recent paper \cite{Storey&Tibshrirani-pnas2003}, Storey and
Tibshirani used a natural cubic spline to fit the data of $c_i$ as
a function of $p_i$ for a given range of p-values, then took the
value of the spline at $p=1$ as the estimation of $N_0/N$. We
compared the Storey-Tibshirani method with Eq.(\ref{eq-cmedian}),
an advantage of the latter is that it is computationally much
simpler than the Storey-Tibshirani method. As can be seen from
Table \ref{tab-summary-N1h-N5k}, both our method and the
Storey-Tibshirani method become more accurate as $N$ and/or
$N_0/N$ increases, and in all the cases our method gives slightly
better results, as indicated by the coefficient of variation.

As for the values of $P_l$ and $P_u$, a practical guidance for
choosing them is to set $P_l$ a value between $0.4$ and $0.5$, and
$P_u$ between $0.9$ and $0.95$. In fact, Eq.(\ref{eq-cmedian})
gives quite robust results with respect to changing the values of
$P_l$ and $P_u$ within the recommended range. For a set of
simulation tests with true null fraction $0.8$, using
$(P_l,P_u)=(0.4,0.95)$ gives $c_{mid}=0.800\pm 0.023$, while using
$(P_l,P_u)=(0.5,0.9)$ gives $c_{mid}=0.800\pm 0.024$.

The method here to estimate $N_0/N$ does not depend on the
specific form of statistical tests being used, as long as the
p-values pertaining to the tests are obtained. But similar to the
BUM method and the the Storey-Tibshirani method, the method we are
proposing here also implicitly assumes that the multiple test
statistics are independent, or at least the true null statistics
are independent. In the context of microarray experiments, this
would require that the null genes' expressions are independent of
each other. This may be not realistic, thus the estimation of the
fraction of null genes based on these methods will be less
accurate. An extreme example is when all the null genes in each
biological sample behave in a concerted manner, and all the
non-null genes express in a synchronized way, then the p-values we
observe will be concentrated on two separate points, one for all
the null genes and one for the non-null genes. Such a situation
will defy all the methods for estimating $N_0/N$ discussed here.
Estimating the fraction of null genes with possibly strong
inter-gene dependence is an important issue, and probably a very
difficult one, especially without specifying their structure of
interdependence beforehand. This is beyond the scope of current
study, and is an issue worth of future investigation and
continuous efforts. Until further statistical advances are made in
this respect, the method we proposed in this paper can serve as an
approximation for estimating the fraction of null statistics.

\section{Discussion}
The data volume generated by microarray studies combined with the
intrinsic variability of the system demands that rigorous
statistical analysis be applied to the data to avoid the problem
of false positives and/or low successful rate in DGE detection. In
this study we have taken into account all the major variables
associated with microarray data. The procedure proposed in this
paper deals with fluorescent label bias often present in
microarray experiments. A t statistic has been derived for
hypothesis testing based on a model that describes each gene
individually with its own set of parameters. An advantage of this
design is that if there exists any fluorescent biases ($D_{g}\neq
D_{r}$) for some genes they will be corrected by the reverse
labelling procedure. For genes with no fluorescent bias (for
example, some genes may have $D_{g}= D_{r}$) the method will
perform equally satisfactorily.

In this work, we have adopted the normality assumption, which
leads to the test statistic $t_0$ following the Students' t
distribution under the null hypothesis. Thus the successful
detection rate $S$ can be calculated in closed form. While the
normality assumption seems reasonable with common technologies,
especially for the measurement error $\epsilon _{visc}$, large
scale replicate experiments have not yet been performed to make a
precise assessment \cite{Baldi&Hatfield2002}. If normality is not
met, $R$ defined in Eq.(\ref{eq-R}) will continue to be an
unbiased estimator of the quantity of interest but $t_0$ will not
follow Students' t distribution. In this case some non-parametric
methods
\cite{Efron-etal-2001,Tusher-etal-pnas2001,Pan-etal-2001,Pan-bioinfo2003}
could be employed. While those methods can be readily applied to
microarrays with a common reference design, where the systematic
dye bias subtracts out in the calculation of the test statistic,
the application of those methods to the direct comparison design
needs to be further developed and investigated. If non-parametric
methods have to be used the rate of successful detection cannot be
as readily calculated as in Eq.(\ref{eq-S}).

In the published literature it is a common practice to apply some
form of normalization (global or local) to remove systematic
biases before the statistical analysis of microarray data. Here we
are proposing to remove much of the systematic bias by
experimental means, i.e. by a dye-swapping procedure. Since the
model deals with the fluorescent bias for each gene individually,
no other local normalization procedure (e.g. LOWESS
\cite{Quackenbush2002}) should be applied before the statistical
testing procedure given here. However, some form of global
normalization is appropriate, such as that utilized by Pollack
{\em et al} \cite{Pollack1999}, or that described in
\cite{Gant&Zhang2002}, where the log-ratios in a microarray
dataset are globally shifted so that the most probable value of
log-ratio becomes $0$. The purpose of global normalization is to
adjust the effect of global factors that could generally affect
the fluorescence, such as a difference between the overall
concentrations of two mRNAs, and possibly the difference of
photo-amplifier voltages used between the two fluorescent channels
when the microarray image was scanned. All the local
feature-specific bias is looked after by the reverse labelling and
statistical testing procedure proposed here.

Finally a word for the overworked bench researcher facing the
prospect of multiple hybridizations in order to achieve a
reasonably high level of $S$ without having to contend with an
unsatisfactory false positive rate. What can be regarded as
reasonable? This depends on the desired outcome of the experiment.
If for example the interest is in defining genes which might give
rise to differential susceptibility, then there will be a desire
to have a high value of $S$ in order not to miss any potential
candidate genes. There would be two ways of achieving this, either
by increasing the number of hybridizations or by accepting a
higher false positive rate. In an experiment such as the one
described then the candidate genes will probably be verified by
other methods downstream. Therefore the balance is driven by the
need to achieve a high $S$ and the decision is between whether it
is more economical to use more microarrays, or put more resource
into downstream verification. Where no downstream verification of
DGEs identified in a microarray experiment are proposed then it is
essential to maintain a low value of false positive rate, at the
expense of $S$ if the total number of microarrays is limiting.
This study does not seek to put a figure on the number of
microarrays that should be hybridized in an experiment. Rather a
framework is provided for the experiment designer to decide on the
number of microarrays to hybridize taking into account the system,
availability of sample, downstream analysis primarily and the
objective of the experiment.

\section*{Acknowledgement}
We wish to acknowledge the support of the microarray team of the
MRC Toxicology Unit particularly Reginald Davies, David J. Judah,
JinLi Luo and Joan Riley. We thank Andy Smith and Michael Festing
for critical readings of the manuscript and helpful discussions.
We also thank anonymous referees for very helpful and constructive
comments.

\bibliography{bioinfo}
\bibliographystyle{apsrev}
%\bibliographystyle{achicago}
%\bibliographystyle{oxford_en}
%\bibliographystyle{alpha}
%\bibliographystyle{pnas}

%\end{document}

\begin{table*}
\caption{\label{tab-summary-N1h-N5k} The fraction of null genes as
estimated by Eq.(\ref{eq-cmedian}) ($c_{mid}$) and by the
Storey--Tibshirani method ($\pi_0$). Parameters used are: $\mu=1$,
$\sigma_T=0.5$,$n_f=2$, $n_r=2$, $P_l=0.4$, $P_u=0.95$. Results
are based on $16$ simulations for each cell in the table. cv, the
coefficient of variation, is defined as the standard deviation
divided by the true value of null fraction, $N_0/N$.}
%\begin{ruledtabular}
\begin{tabular}{cc|ccc|ccc|ccc|ccc}
\hline
&&\multicolumn{3}{c|}{$N=100$}&\multicolumn{3}{c|}{$N=500$}
&\multicolumn{3}{c|}{$N=1000$}&\multicolumn{3}{c}{$N=5000$}\\
$N_0/N$&
&mean&stdev&cv&mean&stdev&cv&mean&stdev&cv&mean&stdev&cv\\
\hline
0.2&$c_{mid}$&0.186&0.040&0.200&0.205&0.017&0.085&0.197&0.013&0.067&0.201&0.009&0.044\\
0.2&$\pi_0$&0.158&0.109&0.544&0.209&0.061&0.307&0.178&0.042&0.212&0.203&0.017&0.087\\
\hline
0.8&$c_{mid}$&0.767&0.112&0.140&0.807&0.047&0.059&0.805&0.031&0.038&0.800&0.023&0.029\\
0.8&$\pi_0$&0.724&0.284&0.355&0.785&0.097&0.121&0.792&0.064&0.080&0.807&0.060&0.075
\\ \hline
\end{tabular}
%\end{ruledtabular}
\end{table*}

\end{document}